\documentclass{PoS}

\usepackage{epsfig}

\usepackage{array}

\usepackage{amsmath}

\usepackage{bm}

\newcommand{\be}{\begin{equation}}
\newcommand{\ee}{\end{equation}}

\newcommand{\beq}[1] {\begin{equation}\label{#1} }
\newcommand{\eeq} {\end{equation} }
\newcommand{\bea}[1]{\begin{eqnarray}\label{#1} }
\newcommand{\eea}{\end{eqnarray}}

\def\beqn{\begin{eqnarray}}
\def\eeqn{\end{eqnarray}}
\def\beq{\begin{equation}}
\def\eeq{\end{equation}}
\def\bea{\begin{equation}}
\def\eea{\end{equation}}

\def\ol{\overline}

\title{Radiative Generation of Neutrino Masses and its Experimental Signals}

\ShortTitle{Radiative neutrino masses}

\author{\speaker{K.S. Babu}\thanks{Work is supported in part by the US Department of Energy Grants
DE-FG02-04ER41306 and DE-FG02-ER46140;  OSU-HEP-11-2}\\
        Department of Physics, Oklahoma State University, Stillwater, OK 74078, USA\\
        E-mail: \email{babu@okstate.edu}}

\abstract{Tiny neutrino masses can arise naturally via loop diagrams.
After a brief review of the radiative mass generation mechanism, I
present a new model wherein TeV scale leptoquark scalars induce neutrino masses
via two--loop diagrams.  This model predicts the neutrino oscillation parameter
$\sin^2\theta_{13}$ to be close to the current experimental limit.
The leptoquarks are accessible to experiments at the LHC since their masses must lie below 1.5 TeV,
and their decay branching ratios probe neutrino oscillation parameters.
Rare lepton flavor violating
processes mediated by leptoquarks have an interesting pattern: $\mu
\rightarrow e \gamma$ may be suppressed, while $\mu \rightarrow 3 e$
and $\mu-e$ conversion in nuclei are within reach of the next
generation experiments.  Muon $g-2$ receives new positive contributions, which can
resolve the discrepancy between theory and experiment.
New CP violating contributions to $B_s-
\ol{B}_s$ mixing via leptoquark box diagrams are in a range that can
explain the recently reported dimuon anomaly by the D\O\ collaboration.}

\FullConference{35th International Conference of High Energy Physics - ICHEP2010,\\
		July 22-28, 2010\\
		Paris France}

\begin{document}
\section*{}
\vspace*{-0.5in}

The standard paradigm for explaining tiny neutrino masses is the seesaw mechanism,
which generates an effective dimension--5 operator
${\cal O}_1 = (L L HH)/M$, suppressed by the mass scale $M$ of the
heavy right--handed neutrino.  ($L$ here denotes lepton doublets, while $H$ is the
Higgs doublet.) Oscillation data suggests that in
this scenario $M \sim 10^{14}$ GeV, which is well beyond the reach of
foreseeable experiments for direct scrutiny.
An interesting alternative to the high scale seesaw mechanism is
radiative mass generation.  The smallness of
neutrino masses can be understood as originating from loop and
chirality suppression factors.  The scale of new physics can naturally be around a TeV
in this scenario.  The simplest among this class of
models is the Zee model \cite{zee1} where neutrino masses
are induced as one--loop radiative corrections arising from the
exchange of charged scalar bosons.  The effective lepton number violating operator in this
model is ${\cal O}_2= LLLe^cH/M$.  To convert this operator to neutrino mass, a loop
diagram is necessary, as shown in the first diagram of Fig. 1. Here $\Phi_1$ is a charged scalar singlet transforming as
$(1,1,1)$ of $SU(3)_c \times SU(2)_L \times U(1)_Y$ gauge symmetry and coupling
to lepton doublets as $f_{ij}L_i L_j\Phi_1^+$ with $f^T = -f$.  $\Phi_2 (1,2,-1/2)$ is a second Higgs doublet that generates
charged lepton masses. The cubic scalar coupling $\Phi_1 \Phi_2 H$ in the scalar potential, along with the term $(\Phi_2H)^2$ ensure that lepton number is explicitly broken.  The neutrino mass in this model is given by $m_\nu \sim (fm_\ell^2+ m_\ell^2f^T)/(16 \pi^2 M_\Phi)$, which
for $M_\Phi = 1$ TeV and $f= 10^{-3}$ yields $m_\nu \sim 0.05$ eV, of the right order to explain
atmospheric neutrino oscillations.  The simplest version of the Zee model is however excluded by neutrino oscillation data,
since it predicts all the diagonal entries of the neutrino mass matrix to be zero, which is inconsistent.

In a second class of models, neutrino masses arise as two--loop radiative corrections \cite{zee2}
via the exchange of a singly charged scalar $\Phi_1(1,1,1)$ and a doubly charged scalar $\Phi_2(1,1,-2)$,
as shown in the second diagram of Fig. 1.  The Yukawa couplings $f_{ij}L_i L_j \Phi_1^+ + g_{ij}e^c_i e^c_j \Phi_2^{--}$, with $f^T=-f,~
g^T = g$,
along with the cubic scalar coupling $\Phi_1^+ \Phi_1^+ \Phi_2^{--}$ ensure lepton number violation.
The effective operator of this model is ${\cal O}_9=LLL
e^c L e^c/M^2$, which requires two--loop dressing to convert to neutrino mass. Since $m_\nu \sim (f m_\ell g m_\ell f^T)/[(16\pi^2)^2 M_\Phi]$
in this model, for $f\sim g\sim 0.1$, and $M_\Phi \sim 1$ TeV, $m_\nu \sim 0.05$ eV is generated.
This model is consistent with neutrino oscillation
data, and predicts the lightest neutrino to be nearly massless.  Phenomenology of this model has been studied in Ref.
\cite{babu1}.  The cross section for the production of a 1 TeV $\Phi_2^{--}$ at the LHC  ($\sqrt{s} = 14$ TeV) is about
20 $fb$, which should be observable with its decay into same sign dileptons.
\begin{figure}[h]
\centering
      \includegraphics[scale=0.4]{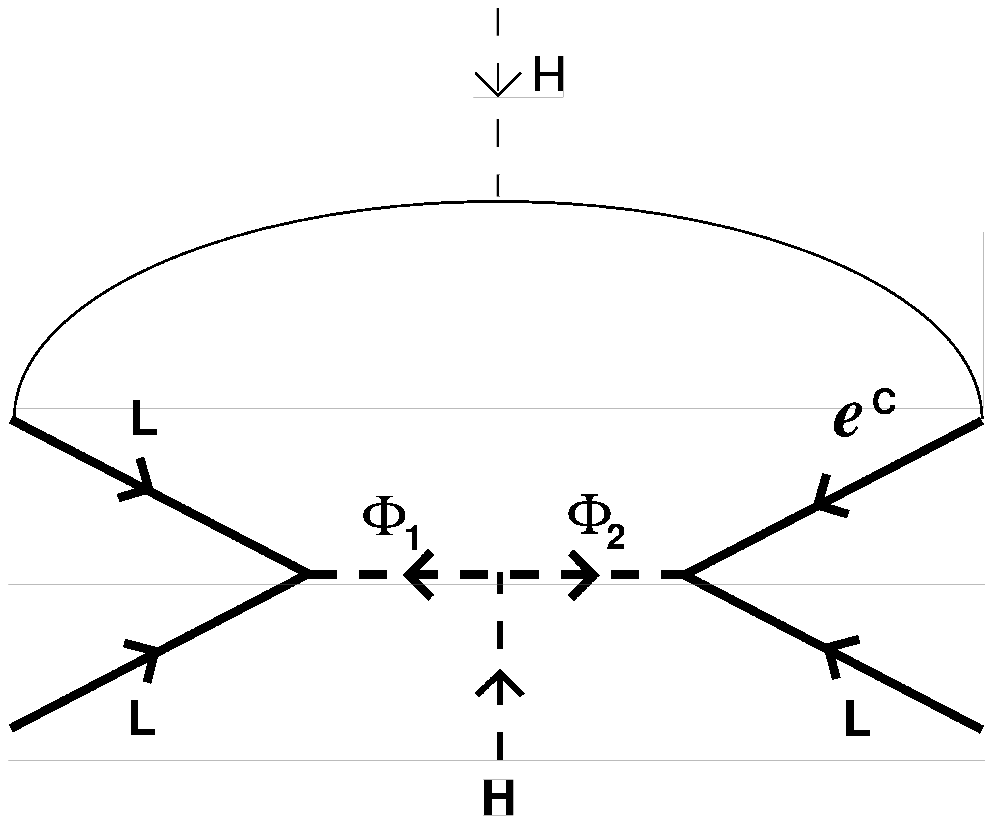}
      \hspace*{2cm}
        \includegraphics[scale=0.35]{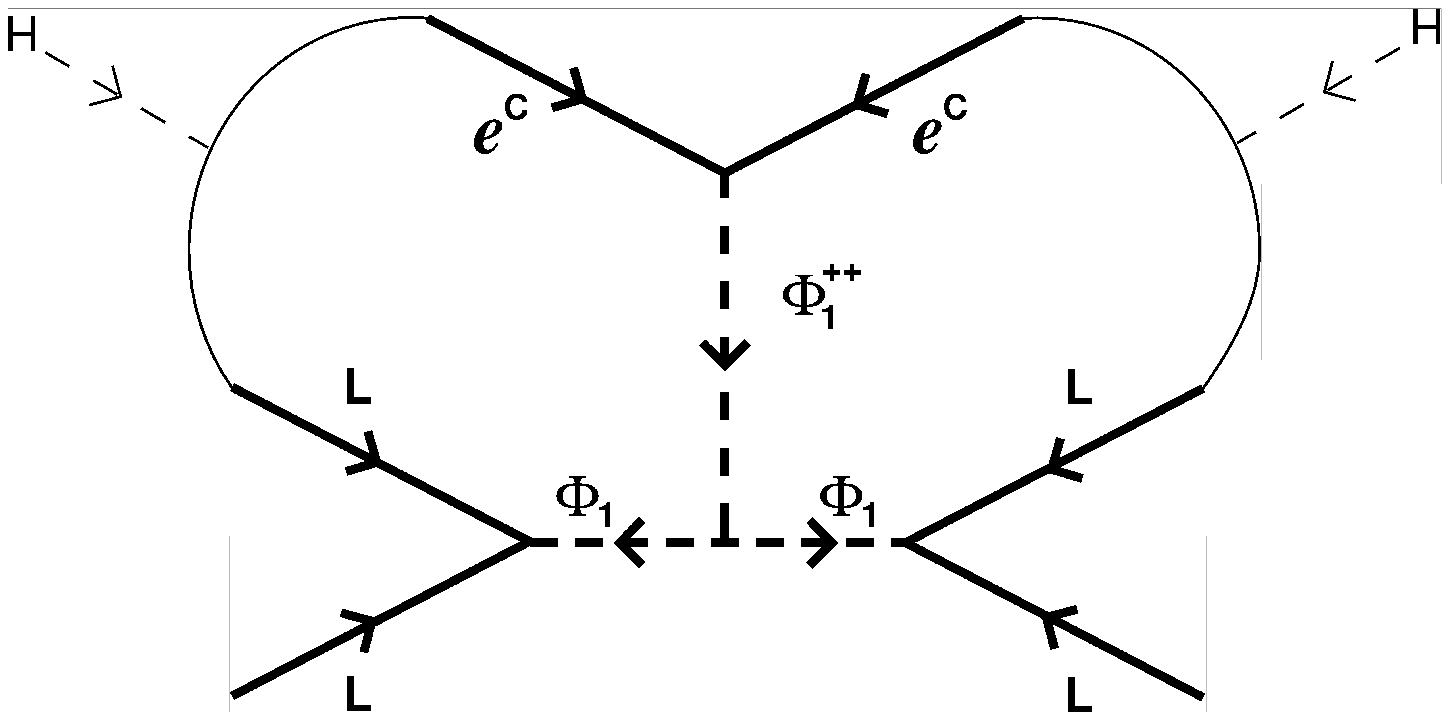}
    \caption{\footnotesize Loop diagrams generating small neutrino masses in the Zee model (left) and in the model of Ref. \cite{zee2} (right).}
    \label{double-beta}
\end{figure}

\begin{figure}[h]
\centering
    \includegraphics[scale=0.4]{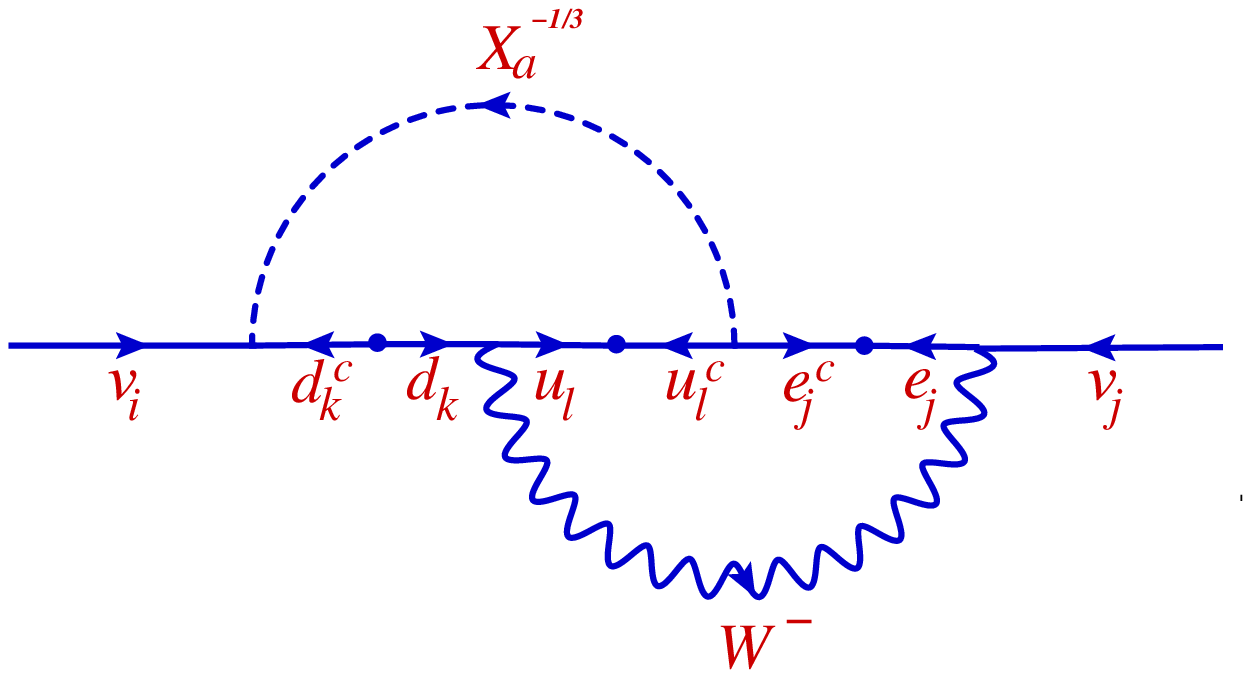}
    \includegraphics[scale=0.42]{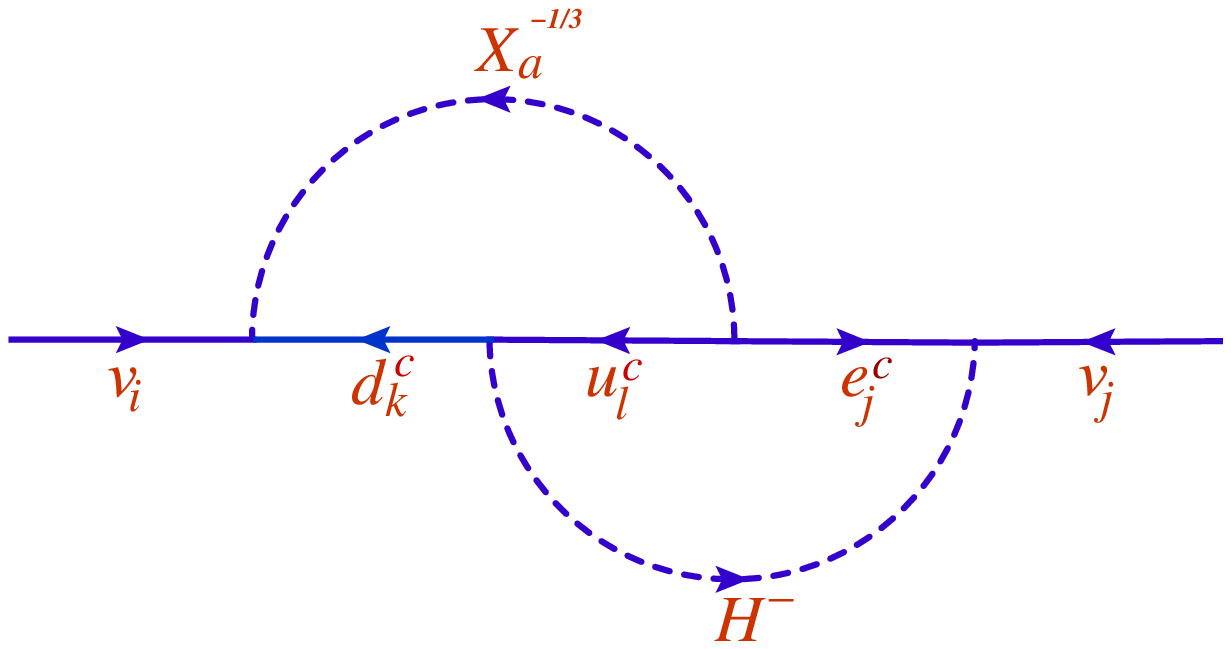}
    \caption{\footnotesize Two--loop diagrams contributing to neutrino mass generation via ${\cal O}_8$.}
    \label{two--loop}
\end{figure}

A classification of low-dimensional effective $\Delta L = 2$ lepton
number violating operators that can lead to neutrino masses has been
given in Ref. \cite{babu-leung}.  The list of operators includes ${\cal O}_3 = L^i L^j Q^kd^cH^l \epsilon_{ik} \epsilon_{jl}$, which
appears in the context of $R$--parity violating supersymmetry.  The operator
${\cal O}_8 = L_i \overline{e^c} ~\overline{u^c} d^c H_j \epsilon^{ij}$
is the subject for the remainder of this paper, which leads to an interesting neutrino mass model \cite{bj}.
${\cal O}_8$ is most directly induced by the exchange of scalar
leptoquarks (LQ).  The order of magnitude of $m_\nu$ arising from ${\cal O}_8$ is $m_\nu \sim ({m_t m_b m_\tau \mu v})/[{(16 \pi^2)^2 M_{\rm LQ}^4}]$,
where $\mu$ is the coefficient of a cubic scalar coupling, and $v= 174$ GeV
 is the electroweak VEV. In order to generate $m_\nu \sim 0.05$ eV, it
 is clear that $M_{\rm LQ}$ must be of order TeV, which would be within reach of the LHC. The scalar sector consists of the
leptoquark multiplets  $\Omega(3,2,1/3) \equiv (\omega^{2/3},\,\omega^{-1/3})$ and  $\chi^{-1/3} (3,1,-2/3)$.
Assuming global baryon number conservation, the Lagrangian relevant for neutrino mass is
\begin{eqnarray}
{\cal L}_{\nu} &=& Y_{ij} ( \nu_i d^c_j \omega^{-1/3} - \ell_i d_j^c
\omega^{2/3}) + F_{ij} \ell_i^c u_j^c \chi^{-1/3} -\mu(\omega^{-2/3}
H^+ + \omega^{1/3} H^0) \chi^{-1/3} + {\rm h.c.} \label{Lag}
\end{eqnarray}
The cubic scalar coupling will generate mixing between $\omega^{-1/3}$ and
$\chi^{-1/3}$, we denote the mass eigenstates $X^a$,
their masses $M_{1,2}$, and the mixing angle $\theta$.  Neutrino masses are induced via Fig. \ref{two--loop}.

Combining constraints from flavor changing processes, we find the neutrino mass matrix to be
\begin{eqnarray}\label{Mnu2}
M_\nu ~\simeq~ m_0 \left( \begin{array}{ccc} 0 & \frac{1} {2} \frac{m_\mu}
{m_\tau} x y & \frac{1}{2} y \cr \frac{1}{2} \frac{m_\mu}{m_\tau} x y & \frac{m_\mu}{m_\tau} x z & \frac{1}{2} z + \frac{1}{2}
\frac{m_\mu}{m_\tau}x \cr \frac{1}{2} y & \frac{1}{2} z + \frac{1}{2} \frac{m_\mu}{m_\tau}x & 1+w\end{array} \right).
\end{eqnarray}
Here
$x = F^*_{23}/F^*_{33}, \,y = Y_{13}/Y_{33},~ z =Y_{23}/Y_{33}$,
$w = (F_{32}^*/F_{33}^*)(Y_{32}/Y_{33})(m_c/m_t)(m_s/m_b)
(I_{jk2}/I_{jk3})$ and  $m_0 = (6 g^2 \sin 2\theta F^*_{33} Y_{33}~ I_{jk3})(m_t m_b m_\tau)/[(16\pi^2)^2M_1^2]$.
$I_{jk3}$ denotes the two loop integral function shown in Fig. 3, with the internal up--type quark being the top.
\begin{figure}[h]
\centering
    \includegraphics[height=3cm]{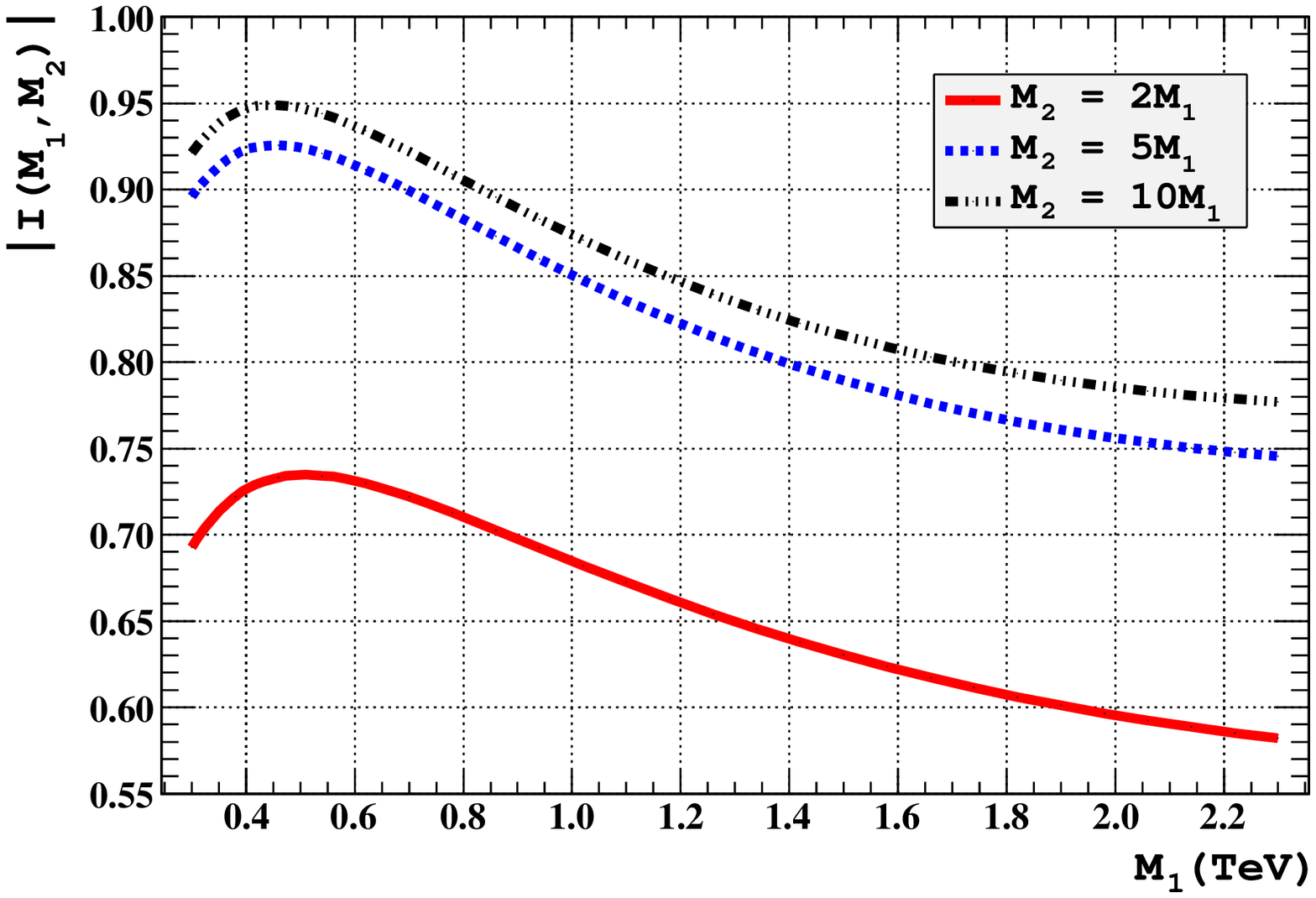}
    \includegraphics[height=3cm]{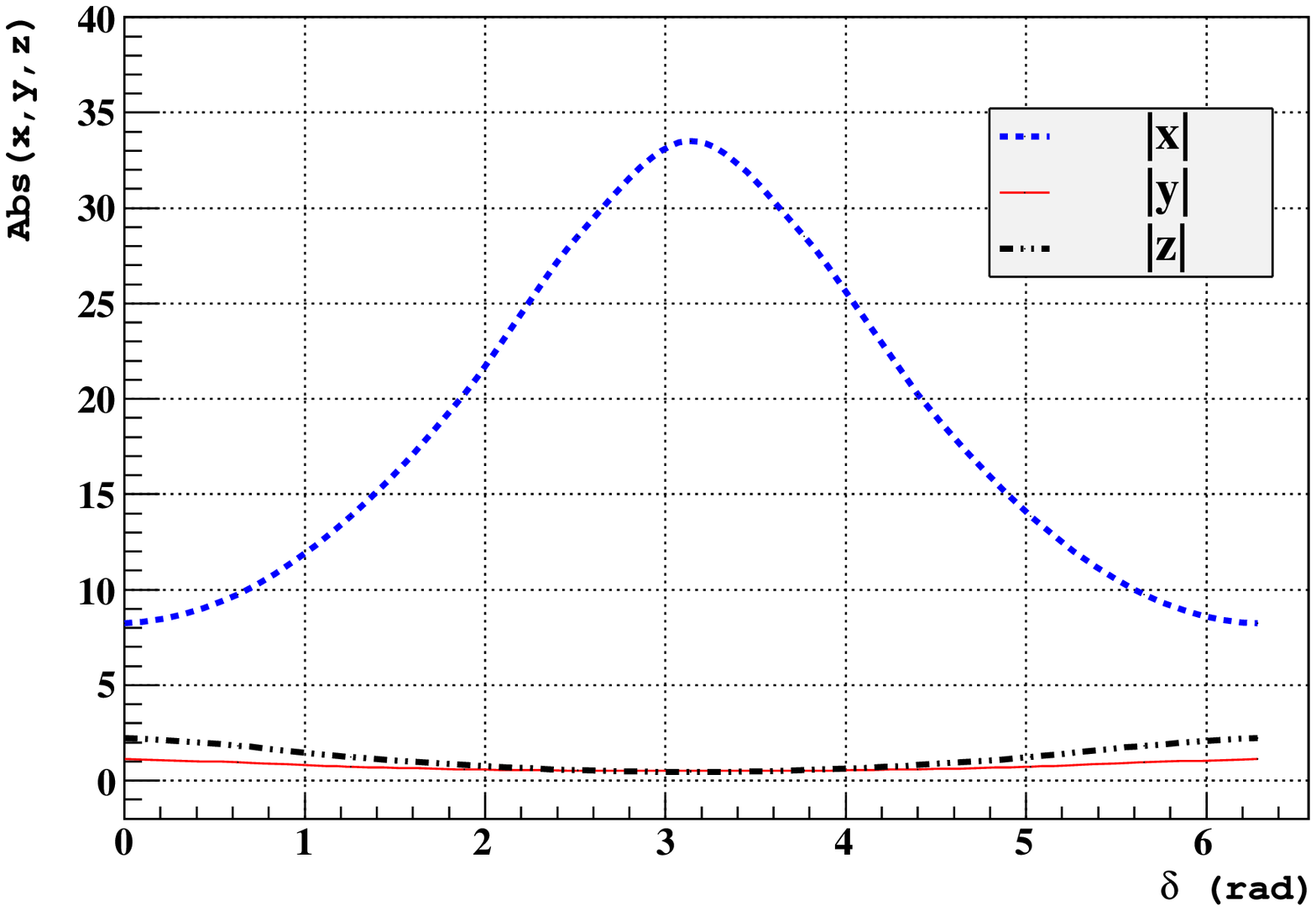}
    \includegraphics[height=2.9cm]{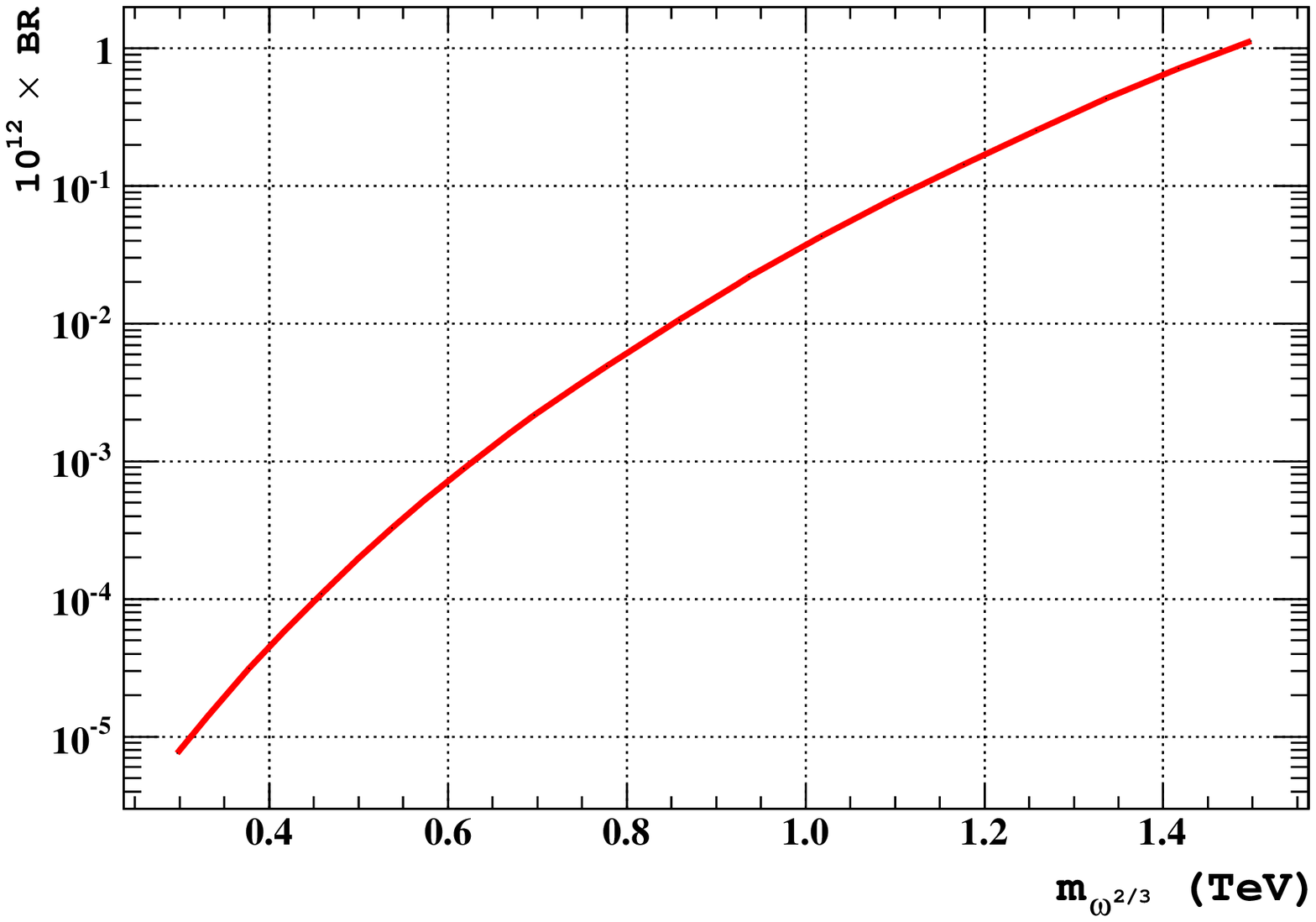}
    \caption{\footnotesize The integral $I_{jk3}$ versus $M_1$ (left); $|x|,\,|y|,\,|z|$  versus $\delta$ (middle);
    $\mu \rightarrow 3e$ branching ratio versus $m_{\omega^{2/3}}$ (right).}
    \label{loop-integral}
\end{figure}
Since the (1,1) entry is zero, and $w$ is highly suppressed, the determinant of $M_\nu$ is nearly zero. This leads to
the predictions $m_1 \simeq 0$, and $\tan^2\theta_{13} \simeq {m_2/m_3}
\sin^2\theta_{12}$ in the standard parametrization of neutrino mixing.  This leads to  $\sin^2\theta_{13} = (0.044-0.051)$,
which is near the current limit.  A consistent fit to global oscillation parameters is obtained.
The parameters $x,\,y,\,z$ for such a fit are
plotted as functions of the unknown CP violating phase $\delta$ in Fig. 3 (middle panel).
We see that  $|x|\gg1$ and $|y|,~|z| \sim 1$.  These values fix the branching ratios
of the leptoquarks:
$\Gamma (\omega^{2/3} \rightarrow e^+ b):~ \Gamma (\omega^{2/3}
\rightarrow \mu^+ b):~ \Gamma (\omega^{2/3} \rightarrow \tau^+ b)=
|y|^2:~ |z|^2:~1$,  and $\Gamma (X_a^{-1/3} \rightarrow \mu^- t):~ \Gamma (X_a^{-1/3}
\rightarrow \tau^- t) =
 |x|^2~: ~1$.  Measuring these decays will thus probe CP violation
in neutrino oscillations.  From the experimental limit on $\mu \to 3e$ and $\mu-e$ conversion in nuclei, we derive the upper limit on
$|Y_{13}^*Y_{23}|$ as a function of $\omega^{2/3}$ mass. Since
neutrino oscillation data requires $Y_{i3}$ to be of the same order for
$i=1-3$, one can also determine an upper limit on $Y_{33}$.
Combining these, we obtain an upper limit of 1.5 TeV on
$M_1$, as shown in Fig. 3  as a function of $\mu \rightarrow 3e$ branching ratio (right panel).

Since the leptoquarks of the model are at the TeV scale, they can mediate lepton flavor violation
through the $Y_{ij}$ and $F_{ij}$ couplings.  Experimental limits on these couplings have been
satisfied in our neutrino fit.  The processes we consider are
$\mu^- \to e^- \gamma$, $\mu^- \to e^+e^-e^-$, $\mu-e$ conversion in
nuclei, $\tau^- \to e^-\eta$, $\tau^- \to \mu^- \eta$,
$B_{s,d}-\ol{B}_{s,d}$ mixing, $K-\ol{K}$ mixing, $D-\ol{D}$ mixing,
$D_s^\pm \rightarrow \ell^\pm \nu$ decay, muon $g-2$, $\pi^+\to
\mu^+ \ol{\nu_e}$ decay, and neutrinoless double beta decay.
Interestingly, we find that $\mu \rightarrow e \gamma$ mediated by the $\omega$ LQ is suppressed by a GIM--like mechanism.
For the $\chi$ LQ mass of 1 TeV, we obtain from $\mu \rightarrow e \gamma$, $\left|\sum_i F_{1i}^*F_{2i}\right| \leq 6.7 \times 10^{-3}$,
but no constraint for the $\omega$ LQ.  If all the LQ masses are 1 TeV, we obtain from $\mu \rightarrow 3e$ the limits
$|Y_{13}Y_{23}| < 7.6 \times 10^{-3}$, $|F_{13}F_{23}| < 1.8 \times 10^{-3}$, and from $\mu-e$ conversion in nuclei slightly better
limits $|Y_{13}Y_{23}| < 4.6 \times 10^{-3}$, $|F_{13}F_{23}| < 1.9 \times 10^{-4}$.  The decays $\tau^- \rightarrow e^- \eta$ and
$\tau^- \rightarrow \mu^- \eta$ provide the limits $|Y_{12} Y_{32}| < 1.2 \times 10^{-2}$, $|Y_{22}Y_{32}| < 1.0 \times 10^{-2}$,
now for the $\omega$ LQ mass of 300 GeV.  $g-2$ of the muon receives new positive contributions from the $\chi$ LQ, which can
be as large as $\delta(g-2) \approx 12 \times 10^{-10}$ for $\chi$ mass of 300 GeV.  This will be nicely consistent with the
indication that $\delta(g-2)= (24.6 \pm 8.0)\times 10^{-10}$.

Recently the D\O\ Collaboration has reported a 3.2 sigma excess in the
like-sign dimuon asymmetry compared to theory. A likely explanation is that there is a new source
of CP violation in $B_s-\overline{B}_s$ mixing, which can arise from
leptoquark box diagrams.  If the $B_s-\overline{B}_s$ mass difference
is written as $\Delta M_s = \Delta M_s^{\rm SM} \left|1+h_se^{2i\sigma_s}\right|$, then a good fit to the data is obtained
for \cite{ligeti} $\{h_s \sim 0.5,~ \sigma_s \sim
120^o\}$ or \{$h_s \sim 1.8,~\sigma_s \sim 100^o$\}.  $h_s \sim 0.5$ is realized in our model
for $|Y_{32}| \sim 1$ and $m_\omega = 390~{\rm GeV}$ or $|Y_{32}| \sim 0.77$ and $m_\omega = 300~{\rm GeV}$. The phase $\sigma_s$ is unconstrained.
This scenario will predict the branching ratio for $B_s \rightarrow \tau^+ \tau^-$ at the 0.25\% level,
compared to the standard model value of $10^{-6}$.
Finally, neutrinoless double beta decay proportional to neutrino mass is suppressed in this model. However,
it can proceed via the vector--scalar exchange
process \cite{vector-scalar}.  The diagram involves exchange of one leptoquark and one $W$ boson.  We obtain the constraint
$\left|Y_{11}^*F_{11}\right| < 1.7 \times 10^{-6}\left(\frac{M_1}{1~
{\rm TeV}}\right)^2\left(\frac{{\rm 0.5~TeV}}{\mu}\right)$ from this process, indicating that neutrinoless double
beta decay may be observable, in spite of the mass hierarchy being normal.

\end{document}